\documentclass{article}

\usepackage[final]{neurips_2024}

\usepackage[utf8]{inputenc} % allow utf-8 input
\usepackage[T1]{fontenc}    % use 8-bit T1 fonts
\usepackage{hyperref}       % hyperlinks
\usepackage{url}            % simple URL typesetting
\usepackage{booktabs}       % professional-quality tables
\usepackage{amsfonts}       % blackboard math symbols
\usepackage{nicefrac}       % compact symbols for 1/2, etc.
\usepackage{microtype}      % microtypography
\usepackage{multirow}
\usepackage{enumitem}
\setlist[enumerate]{nosep}
\usepackage{bbm}
\usepackage{graphicx}
\usepackage{makecell}
\usepackage[table]{xcolor} 

\usepackage{amsmath}
\usepackage{amssymb}
\usepackage{mathtools}
\usepackage{amsthm}
\usepackage{thmtools}

\usepackage{tabularx}
\usepackage{caption}

\theoremstyle{plain}

\theoremstyle{definition}

\theoremstyle{remark}

\newcommand{\ours}{\textsc{Ours}}

\newcommand{\blue}[1]{\textcolor{blue}{#1}}
\newcommand{\todo}[1]{%
    \ifnum#1>0%
        \blue{todo}, \todo{\numexpr#1-1\relax}%
    \fi%
}
\definecolor{verylightgray}{gray}{0.1}  
\definecolor{lightblue}{RGB}{220, 240, 255}  

\newcommand{\ms}[2]{\scalebox{0.95}{{#1}}\scalebox{0.85}{\tiny{}$\pm${#2}}}
\newcommand{\mstop}[2]{\cellcolor{lightblue}{\bf \scalebox{0.95}{{#1}}}\scalebox{0.85}{\tiny{}$\pm${#2}}}

\newcommand{\m}[1]{\scalebox{1}{{#1}}}
\newcommand{\mtop}[1]{\cellcolor{lightblue}{\bf \scalebox{1}{{#1}}}}

\title{Learning to refine domain knowledge\\
for biological network inference}

\author{%
  Peiwen Li \\
  Shenzhen International Graduate School\\
  Tsinghua University\\
  \texttt{lpw22\{at\}mails.tsinghua.edu.cn} \\
  \And
  Menghua Wu \\
  Department of Computer Science \\
  Massachusetts Institute of Technology \\
  \texttt{rmwu\{at\}mit.edu} \\
}

\begin{document}

\maketitle

\begin{abstract}
Perturbation experiments allow biologists to discover causal relationships between variables of interest, but the sparsity and high dimensionality of these data pose significant challenges for causal structure learning algorithms.
Biological knowledge graphs can bootstrap the inference of causal structures in these situations, but since they compile vastly diverse information, they can bias predictions towards well-studied systems.
Alternatively, amortized causal structure learning algorithms encode inductive biases through data simulation and train supervised models to recapitulate these synthetic graphs.
However, realistically simulating biology is arguably even harder than understanding a specific system.
In this work, we take inspiration from both strategies and propose an amortized algorithm for refining domain knowledge, based on data observations.
On real and synthetic datasets, we show that our approach outperforms baselines in recovering ground truth causal graphs and identifying errors in the prior knowledge with limited interventional data.
\end{abstract}

\section{Introduction}
\label{intro}

Large-scale perturbation experiments have the potential to uncover extensive causal relationships between biomolecules~\citep{replogle}, which may facilitate myriad applications in drug discovery, from disease understanding to mechanism of action elucidation~\citep{moa}.
Causal structure learning (discovery) algorithms are designed to extract these very relationships directly from data~\citep{causation-prediction-search}.
Yet due to the high number of variables (genes), compounded with the low numbers of observations (cells) per setting (perturbation)~\citep{nadig}, these algorithms struggle to scale and perform robustly on such datasets.
A key challenge is that causal discovery algorithms must not only infer the causal direction between variables, but also which variables are related in the first place.
The latter can be alleviated in part by incorporating noisy priors regarding the data, e.g. by initializing the graph prediction using a biological knowledge graph~\citep{go}.
However, these graphs compile decades of discoveries from disparate experiments, rendering their relevance and correctness uncertain in individual cellular contexts.
While the choice and quality of these priors generally does not impact consistency in the infinite data limit~\citep{gies,bacadi}, there are rarely sufficient data for these guarantees to hold in practice.

An orthogonal line of work aims to capture inductive biases that cannot be easily represented by individual graphs via amortized inference over synthetic data~\citep{ke2022learning,avici}.
A simulator first generates pairs of ``ground truth'' causal graphs and datasets, following known rules regarding the domain of interest.
For example, biological networks have been hypothesized to be scale-free~\citep{sf}, and transcription dynamics can be described through sets of differential equations~\citep{difeq}.
Once these data have been generated, a neural network is trained to map the datasets to their associated causal graphs.
Empirically, the resultant models are more robust to low-data situations in real settings, as they have ideally seen similar, synthetic datasets over the course of training~\citep{sea}.
However, since these approaches incorporate inductive biases via simulation, it is imperative that the simulators accurately reflect the true data.
In reality, it has been observed that truly scale-free graphs are the minority in biology~\citep{sf-rare}, and transcription dynamics can be highly heterogeneous and discontinuous, even within the same cell type~\citep{lenstra2016transcription}.

In this work, we propose an amortized inference framework for \emph{refining} noisy graph priors, to identify causal relationships in low-data regimes.
We are motivated by the idea that while simulating biological data is hard, simulating the types of noise that occur may be easier.
For example, gene-gene relationships may increase/decrease in strength, or appear/disappear based on the cell state, but the directionality of these relationships rarely changes~\citep{dci}.
% Thus, learning to denoise graphs may be much more data-efficient and less biased than biological simulation.
Our model architecture is based on the supervised causal discovery model in \citet{sea}, in which datasets are featurized in terms of local causal graph estimates and summary statistics such as global correlation.
During training, we augment these inputs with a corrupted, undirected graph, sampled at varying levels of noise. The neural network is forced to identify incorrect edges based on data, as well as orient edges where feasible.

We perform extensive experiments on both synthetic datasets and the real-world Sachs proteomics dataset \citep{sachs}, evaluating the model’s ability to 1) predict the ground truth graph and 2) detect errors in noisy priors, particularly under various down-sampling settings to assess performance in low-data regimes. Across all noisy prior conditions and sample sizes, \ours{} consistently achieves strong results in both causal structure learning and error detection in noisy priors. In contrast, baseline methods are significantly affected by the quality of prior knowledge and the amount of data.
% This highlights the robustness and superiority of our approach, particularly in challenging low-data environments and under different noise levels in the priors.
We conclude that high-quality graph priors provide strong starting points for inferring causal relationships, especially when limited data are available, and learning to denoise these priors is more data-efficient than using them to initialize graph predictions.
\section{Background and related work}
\label{related}

\begin{figure}[t]
\centering
\includegraphics[width=\linewidth]{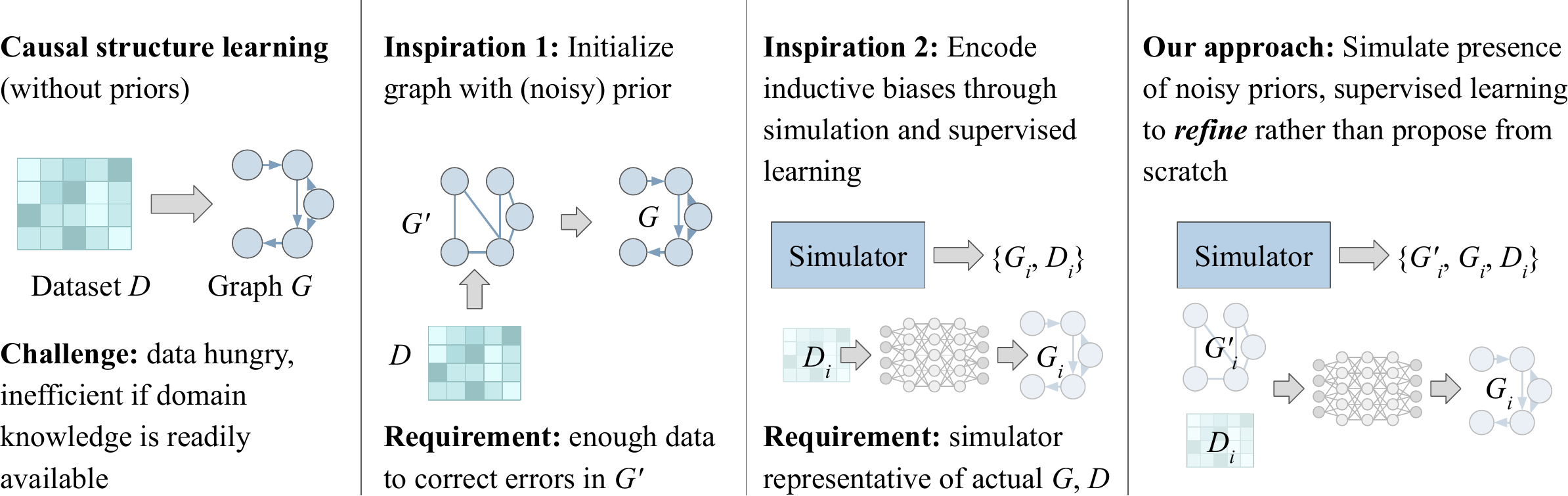}
\caption{Drawing inspiration from amortized causal discovery algorithms, we learn how to refine graph priors, enabling robust graph predictions in low-data regimes.}
\label{fig:overview}
\vspace{-0.1in}
\end{figure}

\subsection{Biological network inference}

Biological network inference is a classic systems biology problem, in which the goal is to uncover interactions between experimentally-quantifiable entities (e.g. genes, proteins) in the form of graphs~\citep{albert2007network,HuynhThu2018GeneRN}. For example, graphs of interest include gene regulatory networks~\citep{grns}, protein-protein interaction networks~\citep{corum}, and metabolic pathways~\citep{reactome}. 
Early efforts towards biological network inference included the DREAM challenges \citep{dream}, which provided harmonized microarray data and were evaluated against known interactions at the time.
% The challenge has motivated the development of numerous algorithms, ranging from statistical approaches to machine learning models, each aiming to improve accuracy in network recovery.
% Gene Regulatory Networks (GRNs) denote interaction structures underpinning patterns of gene expression. Reconstructing these networks has
% been a central effort of the interdisciplinary field of Systems Biology \citep{HuynhThu2018GeneRN}.
% There exist different computational techniques to analyze the causal relationships between the pair of genes and to understand the significance of causal relationship in GRNs \citep{Shambharkar2022BiologicalNG}.
Probabilistic graphical models are commonly used to infer gene regulatory networks in specific disease areas~\citep{mao2004probabilistic, zhao2019cancer, Dai2024GeneRN}.
More recent works have also used graph neural networks to predict ``missing'' edges in these graphs~\citep{Feng2023GeneRN}.
However, while biological network inference algorithms have been applied to a variety of disease areas, many of these methods are typically engineered towards their respective datasets and do not share reproducible code.

\subsection{Causal structure learning}

Causality provides a formal framework for inferring data-generating mechanisms from experimental data.
A causal graphical model is defined by a distribution $P_X$ over a random variables $X$, associated with a directed acyclic graph $G = (V, E)$, where each node $i \in V$ corresponds to a random variable $X_i \in X$, and each edge $(i, j) \in E$ indicates a direct causal relationship from $X_i$ to $X_j$~\citep{causation-prediction-search}. It is common to assume that the data distribution $P_X$ is Markov to $G$, i.e. variables $X_i$ are independent of all other $X_j \not\in X_{\delta_i} \cup X_{\pi_i}$ (not descendants or parents), given its parents $X_{\pi_i}$.
% \begin{equation}
% X_i \perp\!\!\!\perp V \setminus (X_{\delta_i} \cup X_{\pi_i}) \,|\, X_{\pi_i}, \quad \forall i \in V,    
% \end{equation}
% where $\delta_i$ represents the set of descendants of node $i$, and $\pi_i$ denotes the set of parents of $i$.
% Sampling directly from $P_X$ results in \emph{observational} data.
Causal graphical models introduce the concept of interventions on node $i$, by changing the conditional distribution $P(X_i \,|\, X_{\pi_i})$ to a new distribution $\tilde{P}(X_i \,|\, X_{\pi_i})$.
% In this paper, we primarily focus on experiments with perfect interventions, where the dependencies between each node $i$ and its parents are removed by replacing $\tilde{P}(X_i \,|\, X_{\pi_i})$ with $\tilde{P}(X_i)$.

Causal structure learning is the task of predicting causal graph $G$ from dataset $D \sim P_X$.
Classical \textit{discrete optimization methods} operate over the combinatorial space of edge sets, and they make discrete changes to add/delete/orient edges.
These include constraint-based algorithms, such PC and FCI for observational data \citep{fci}, and JCI for mixed data \citep{mooij2020joint}.
While these algorithms can be initialized with an undirected graph as a prior, they cannot recover edges that are not present in this initial skeleton.
There are also score-based methods that optimize a score, which represents the ``goodness'' of a particular graph, with respect to the data. These include GES \citep{ges}, GIES \citep{gies}, CAM \citep{cam}, LiNGAM \citep{lingam} and IGSP \citep{igsp}.
Algorithms like GIES iterate between adding and deleting edges, so they can (in principle) identify edges that are missing from an initial estimate.
However, due to the exponential space of potential graphs and the reliance on statistical power for discrete judgments, these classical approaches scale poorly with the number of variables and require copious data for reliable performance.

On the other hand, \textit{continuous optimization methods} approach causal discovery through constrained continuous optimization over weighted adjacency matrices. Many of these approaches, exemplified by NoTears \citep{notears}, DCDI \citep{dcdi}, and GranDAG \citep{grandag} train a generative model to capture the empirical data distribution, which is parameterized through the adjacency matrix.
% \textbf{Most existing methods rely heavily on large size of data}, particularly a considerable amount of interventional samples, which are often difficult to obtain, in order to achieve a reasonable empirical performance. Also, depending on data availability and the underlying data generation process, causal discovery algorithms may or may not be able to accurately recover the causal graph $G$ in practice.
% \textbf{Applications of causal methods in the field of biomedicine} are also emerging.
Several works have also been specifically designed to address challenges in biological problems. DCD-FG \citep{dcdfg} aims to scale to single-cell transcriptomics data and proposes a low-rank extension of DCDI.
The hybrid IGSP algorithm \citep{igsp} has also been applied to single-cell data.
Prior knowledge can be used to initialize the graph parameters in these frameworks, but the same limitations apply with regards to data-efficiency.

\subsection{Biological knowledge graphs for perturbations}

Knowledge graphs have been indispensable to modeling biological perturbations.
They are commonly used as undirected graphs, over which graph neural networks predict the cellular effects of unseen perturbation~\citep{gears,attentionpert} or infer perturbation targets for active learning~\citep{kexin} and target discovery~\citep{pdgrapher}.
This work focuses on an adjacent but distinct task: of inferring relationships between variables, rather their effects or identity as targets.

\section{Methods}
\label{methods}

\begin{figure}[t]
\centering
\includegraphics[width=\linewidth]{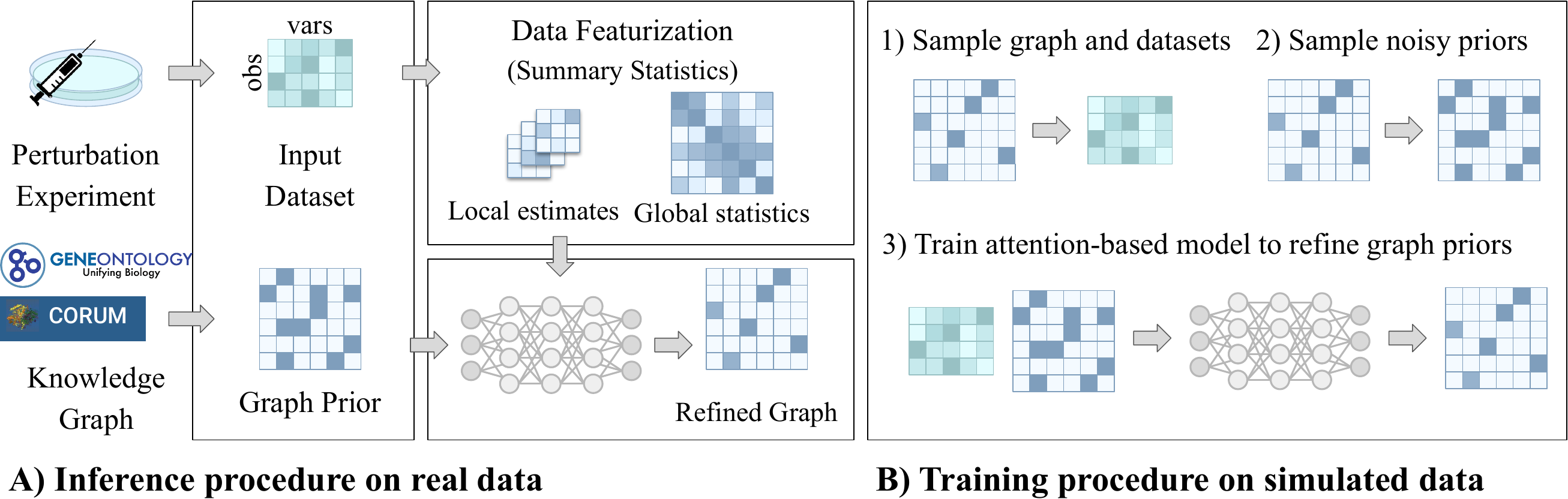}
\caption{A) At inference, we use biological knowledge graphs as noisy graph priors, which we refine with perturbation data.
B) We train an attention-based model to denoise simulated graph priors.}
\label{fig:model}
\vspace{-0.1in}
\end{figure}

Let $D\sim P_X$ be a dataset containing $M$ samples of $N$ variables, and let $G=(V,E)$ be the causal graph that generated $P_X$.
Let $G'=(V, E')$ be an undirected graph, where $E + E^\intercal \approx E'$ but $E + E^\intercal \ne E'$.
Given $D$ and $E'$, the goal is to predict $E$.

\subsection{Inference}

When given a new dataset $D$ and graph prior $E'$, we summarize $D$ in terms of local and global summary statistics, which are combined with $E'$ as input to an attention-based neural network, trained to predict $E$ (Figure~\ref{fig:model}A).
We adapt the Sample, Estimate, Aggregate workflow (\textsc{Sea}, \citet{sea}) to the task of refining noisy graph priors based on data observations as follows.
\begin{enumerate}
    \item (Sample) We sub-sample batches from dataset $D$ to focus on sets of variables that are likely to be related, using heuristics like correlation.
    \item (Estimate) We compute pairwise correlation $\rho$ between all variables as a summary statistic; and run classical causal discovery algorithms over smaller batches, sampled in (1).
    \item (Aggregate) A neural network is provided the $N\times N$ noisy graph estimate $E'$, the $N\times N$ summary statistic, and $T$ estimates of size $k \times k$ (where $k=5$ is small compared to $N$).
    The final output is a $N\times N$ matrix that represents the predicted causal graph, refined from the noisy prior with edges oriented.
\end{enumerate}

\subsection{Training}
\label{sec:training}

At inference time, $E'$ is provided (either a corrupted synthetic graph or a biological knowledge graph).
During training, we sample graph priors by adding noise to the ground truth undirected graph.
We compute $E'$ as follows.
\begin{enumerate}
    \item Sample noise level $p\sim\text{Uniform}(0, 0.5)$,
    and binary mask $M\in\mathbf{1}^{N\times N}$ where
    \begin{equation}
        M_{i,j} = \mathbf{1}\{ z_{i,j} \sim\text{Uniform}(0,1) < p\}.
    \end{equation}
    \item Compute undirected graph $\tilde{E} = E + E^\intercal$.
    \item Compute noisy prior $E'$ where
    \begin{equation}
        E'_{i,j} = \begin{cases}
            \tilde{E}_{i,j} & M_{i,j} = 0 \\
            1 - \tilde{E}_{i,j} & \text{otherwise}.
        \end{cases}
    \end{equation}
\end{enumerate}
We finetune all weights with the binary classification objective of predicting $E$, at the edge level.
Thus, the objective both encourages the model to denoise $E'$ and orient edges.

\subsection{Implementation details}

We adopt the same axial-attention architecture as \textsc{Sea}.
Specifically:
\begin{enumerate}
    \item The $k\times k\times T$ marginal estimates (local graph structures, inferred by standard causal discovery algorithms) are aligned by matching the same edges across estimates, and mapped to a $K\times T \times d$ \emph{marginal} feature, where $K$ is the number of unique edges.
    \item $E'$ is embedding using the same edge embeddings as marginal graphs (since they are in the same input space), and the result is added to that of the global statistic (since they are over the same nodes).
    The result is a $N\times N\times d$ \emph{global} feature.
    % The two $N\times N$ matrices $E'$ and $\rho$ are concatenated and embedded into a $N\times N\times d$ \emph{global} feature.
    \item A series of 2D axial attention layers~\citep{ho2020axial} attend over the rows and columns of both matrices.
    The final output is a $N\times N$ matrix, which is supervised by the (synthetic) ground truth $E$.
\end{enumerate}
Our synthetic training set contains approximately 4000 datasets of size $N=10, 20$ with linear additive and neural network (additive and non-additive) causal mechanisms.
We use pretrained \textsc{Sea} weights with the GIES~\citep{gies} estimation algorithm and inverse covariance as the global statistic.
The introduction of $E'$ does not add any new parameters, as we use the same embeddings as the existing edge estimates, which support undirected edges.
\section{Experimental setup}
\label{sec:experiment-setup}

We evaluate our approach against a variety of causal structure learning baselines on real and synthetic datasets.
While it would be ideal to evaluate entirely on real applications, synthetic experiments allow us to systematically assess how performance changes based on the quality of our noisy priors and the availability of interventional data.
In real settings, it is difficult to quantify the relevance of knowledge graphs to each cell line, disease type, or other factors, as these labels themselves are inherently approximations to the underlying biology.

\subsection{Data preparation}

\paragraph{Biological experiments}

\begin{figure}[t]
    \centering
    \includegraphics[width=\linewidth]{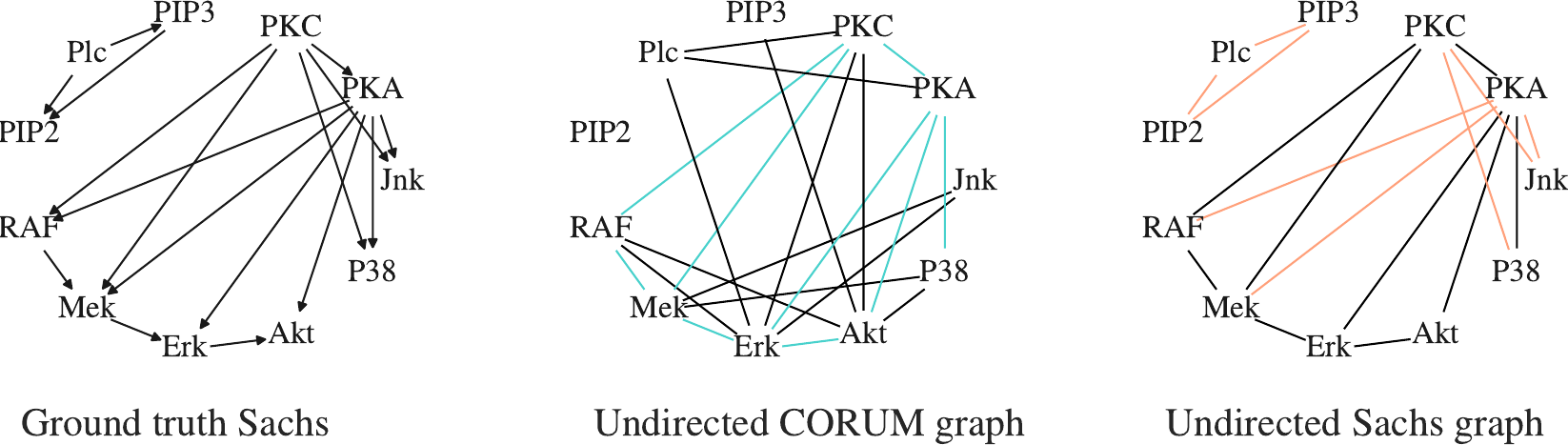}
    \caption{Visualization of ground truth Sachs consensus graph~\citep{sachs} and CORUM knowledge graph~\citep{corum}.
    Blue: Undirected edges present in both CORUM and Sachs.
    Orange: Undirected edges present in Sachs but not CORUM.
    9 of 17 undirected edges in Sachs are present on the CORUM graph; of the 38 pairs of nodes that have no relationship in Sachs, 26 also have no relationship in CORUM.}
    \label{fig:sachs}
    \vspace{-0.1in}
\end{figure}

The Sachs proteomics dataset~\citep{sachs} is a common benchmark for causal structure learning approaches. In this work, we use the subset proposed by \citet{igsp}, which contains 1755 observational samples and 4091 interventional samples, associated with a consensus graph of 11 nodes and 17 edges.
We use both a corrupted version of the ground truth graph (``Synthetic KG'') and the CORUM (comprehensive resource of mammalian protein complexes) knowledge graph~\citep{corum} as noisy priors.
CORUM focuses on physical interactions between proteins, which are likely relevant for, but do not directly translate to quantifiable effects of perturbing certain proteins on others.
Figure~\ref{fig:sachs} depicts the CORUM graph alongside the ground truth.

\paragraph{Synthetic experiments}

% Write this

% Our synthetic data were generated from 
The raw synthetic datasets were generated following \textsc{Dcdi} \citep{dcdi}: (1) sampling Erdős-Rényi graphs with $N = 10, 20$ nodes and $E = N$ expected edges; (2) sampling random instantiations of causal mechanisms (Linear, Neural Networks, Sigmoid with additive Gaussian noise, and Polynomial mechanisms); and (3) iteratively sampling observations in topological order. For each graph, we generated one observational regime and ten interventional regimes, with each regime consisting of $1000$ samples, i.e., $1000\times N$ data points, either entirely observational or following a set of single-node perfect interventions.
% Based on the sampled ground-truth graphs, each containing $N \times N$ relationships, we further introduce noise by randomly selecting $10\%$ and $25\%$ of the edges and reversing their values between $0$ and $1$. Subsequently, we ignore the directionality of the noisy edges to form the final \textit{noisy undirected graphs}.

\paragraph{Down-sampling}

In perturbation experiments, observational data are generally easy to collect and abundant, while interventional data are costly and limited.
For example, large-scale Perturb-seq datasets may include thousands of non-targeting control cells, but only a median of $\sim$50 cells per perturbation~\citep{nadig}.
To emulate this setting on all datasets, we sub-sample 50
or 100 
examples from each interventional regime, while preserving all observational examples.
This allows us to evaluate all models in realistic data settings.

% For both biological and synthetic datasets, to evaluate performance under limited data, which is common in real-world applications, we performed \textit{down-sampling} by selecting either 50 or 100 samples from the 1000 samples in each interventional regime.
% We attempt to verify through down-sampling settings that incorporating domain knowledge helps us perform better with less data.

\begin{table}[t]
% \begin{tabularx}{\linewidth}{XX}

\setlength\tabcolsep{6 pt}
\caption{
Causal discovery and noise detection results on real proteomics dataset~\citep{sachs}, 50 samples per intervention.
The vanilla setting is not evaluated on noise detection since there is no graph prior.
The best results in each category are indicated in \textbf{bold}.
}
\label{table:sachs}
\begin{small}\begin{center}

% \begin{tabular}[b]{c c l cc ccc}
\begin{tabular}[b]{c l cc ccc}
\toprule
\multirow{2}{*}{\makecell{Prior \\ Knowledge}}
% & \multirow{2}{*}{\makecell{Samples per \\ Intervention}}
& Model & \multicolumn{2}{c}{Graph} & \multicolumn{2}{c}{Noise} \\
\cmidrule(l{\tabcolsep}){3-4}
\cmidrule(l{\tabcolsep}){5-6}
& & mAP $\uparrow$ & SHD $\downarrow$ & F1 $\uparrow$ & Acc $\uparrow$ \\
\midrule
\multirow{4}{*}{\makecell{Vanilla}}
% & \multirow{4}{*}{50} 
% & \textsc{Gies} & \mtop{0.21}& \mtop{16} \\
&\textsc{Dcdi-G} & \m{0.14}& \m{23} & --- & --- \\
& \textsc{Dcdi-Dsf} & \m{0.14}& \m{29}& --- & --- \\
& \textsc{BaCaDi} & \m{0.15}& \m{20} & --- & --- \\
& \textsc{Sea} & \mtop{0.20}& \mtop{17}& --- & --- \\

% \cmidrule(l{\tabcolsep}){2-8}
% & \multirow{5}{*}{100} 
% & \textsc{Gies} & \m{0.21}& \m{17} \\
% & &\textsc{Dcdi-G} & \m{0.14}& \m{25}& & & \\
% & & \textsc{Dcdi-Dsf} & \m{0.14}& \m{36}& & --- & \\
% & & \textsc{BaCaDi} & \m{0.15}& \m{24}& & & \\
% && \textsc{Sea} & \mtop{0.29}& \mtop{15}\\

\midrule
\multirow{3}{*}{\makecell{Synthetic KG\\(10\% noise)}}
% & \multirow{3}{*}{50}
& \textsc{Dcdi-G} & \m{0.14}& \m{21} & \m{0.55} & \m{0.73}\\
& \textsc{Dcdi-Dsf} & \m{0.15}& \m{18}& \m{0.53} & \m{0.75}\\
& \ours{} & \mtop{0.25}& \mtop{17}& \mtop{0.73}& \mtop{0.80}\\
% \cmidrule(l{\tabcolsep}){2-8}
% & \multirow{3}{*}{100}
% &\textsc{Dcdi-G} & \m{0.15}& \m{19}& \m{0.57} &\m{0.78} &\m{0.69} \\
% & & \textsc{Dcdi-Dsf} & \m{0.22}& \m{17}& \mtop{0.64}&\m{0.82} &\m{0.75} \\
% \cmidrule(l{\tabcolsep}){4-8}
% & & \ours{} &
% \mtop{0.53}& \mtop{16} &
% \mtop{0.64}&\mtop{0.84} & \mtop{0.77}\\

\midrule\midrule
\multirow{3}{*}{\makecell{Synthetic KG\\(25\% noise)}}
% & \multirow{3}{*}{50}
% & \textsc{Gies} & \m{0.23}& \m{16} & \m{0.80}& \m{0.82}\\
&\textsc{Dcdi-G} & \m{0.19}& \m{19} & \m{0.67} & \m{0.76}\\
& \textsc{Dcdi-Dsf} & \m{0.15}& \m{23}& \m{0.35} & \m{0.60}\\
% & & \ours{} &
% \m{0.48}&\mtop{16} &
% \mtop{0.73}&\mtop{0.84} & \mtop{0.82}\\
& \ours{} & \mtop{0.24}& \mtop{17}& \mtop{0.78}& \mtop{0.80}
\\

% \cmidrule(l{\tabcolsep}){2-8}
% & \multirow{3}{*}{100}
% & \textsc{Gies} & \m{0.23}& \m{16}\\
% &&\textsc{Dcdi-G} & \m{0.14}& \m{20}& \m{0.66}&\m{0.77} &\m{0.74} \\
% & & \textsc{Dcdi-Dsf} & \m{0.15}& \m{24}& \m{0.62}&\m{0.73} &\m{0.67} \\
% % & & \ours{} &
% % \mtop{0.53}& \mtop{16} &
% % \mtop{0.73}&\mtop{0.84} & \mtop{0.82}\\
% && \ours{} & \mtop{0.43}& \mtop{9}\\

\midrule\midrule
\multirow{3}{*}{\makecell{CORUM KG \\ (34\% noise)}}
% & \multirow{3}{*}{50}
% & \textsc{Gies} & \m{0.21}& \m{16} & \m{0.75}& \m{0.82}\\
&\textsc{Dcdi-G} & \m{0.14}& \m{24}& \m{0.44}& \m{0.73}\\
& \textsc{Dcdi-Dsf} & \m{0.15}& \m{20}& \m{0.67}& \mtop{0.82}\\
& \ours{} & \mtop{0.21}& \mtop{16}& \mtop{0.70}& \m{0.78}\\

% \cmidrule(l{\tabcolsep}){2-8}
% & \multirow{3}{*}{100}
% & \textsc{Gies} & \m{0.24}& \m{15}\\
% &&\textsc{Dcdi-G} & \m{0.14}& \m{26}& \m{0.55} &\m{0.71} &\m{0.61} \\
% & & \textsc{Dcdi-Dsf} & \m{0.14}& \m{25}& \m{0.51}&\m{0.69} &\m{0.57} \\
% && \ours{} & \mtop{0.60}& \mtop{11}\\

\bottomrule
\end{tabular}
\end{center}\end{small}
% \end{tabularx}
\vspace{-0.1in}
\end{table}

\subsection{Baselines}

We compare our method to several state-of-the-art causal discovery algorithms, which are able to incorporate prior knowledge by initializing their graph parameters based on these undirected graphs.
Specifically, the ``vanilla'' setting indicates that models do not consider prior knowledge, while the other settings specify the noise level ($10\%, 25\%$) that the undirected graphs were subject to (Section~\ref{sec:training}).
% We compare our methods with both baselines \textsc{Dcdi} \citep{dcdi} that primarily do not incorporate domain knowledge as well as \textsc{BaCaDI}~\citep{bacadi} that do as follows:

\begin{itemize}
    % \item \textsc{Gies}
    % is a classic score-based algorithm for interventional data, which optionally takes a CP-DAG (directed and undirected edges)~\cite{cpdag} to initialize the graph prediction.
    \item \textsc{Dcdi-G} and \textsc{Dcdi-DSF}~\citep{dcdi} are two variations of \textsc{Dcdi}, each employing a different density approximator. \textsc{Dcdi-G} utilizes simple Gaussian distributions, while \textsc{Dcdi-DSF} leverages the more expressive deep sigmoidal flows to represent non-linear causal relationships.
    To incorporate prior knowledge, we initialize the parameters of the Gumbel adjacency matrix such that the initial weighted adjacency reflects to the noisy graph.
    The ``vanilla'' \textsc{Dcdi} initializes these parameters to a matrix of ones, i.e. a fully connected graph.
    % , represented by the Gumbel matrix and is originally a fully connected matrix, corresponds to our noisy graph, which partially reflects knowledge from ground-truth graph.
    % initialize with noisy $G$
    % \item \textsc{Dcdi-DSF}~\citep{dcdi} initialize with noisy $G$
    \item \textsc{BaCaDI}~\citep{bacadi} is a fully Bayesian approach for inferring complete joint posterior over causal structure, parameters of causal mechanisms, and interventions in each experimental context.
    % It inherently consider prior probability and likelihood function beforehand, which serve as the antecedents of inference.
    % prior is noisy $G$
    \item \textsc{Sea}~\citep{sea} is the ``vanilla'' version of our model, which does not incorporate any prior knowledge
    and was not trained to denoise external information.
\end{itemize}

% -------------------------

% Fill this out.
% Baselines that do not incorporate domain knowledge + those that do.

% We should run these 2 (3) at least. Use dcd-fg only when dcdi cannot scale.
% \begin{itemize}
%     \item \textsc{Dcdi}~\citep{dcdi} initialize with noisy $G$
%     \item \textsc{Dcd-Fg}~\citep{dcdfg} initialize with noisy $G$
%     \item \textsc{BaCaDI}~\citep{bacadi} prior is noisy $G$
% \end{itemize}

\subsection{Metrics}

We evaluate our models and all baselines on their ability to 1) predict the ground truth causal graph and 2) detect errors in the prior knowledge.
To assess the quality of the predicted graphs, we report standard causal discovery metrics (discrete and continuous).
\begin{enumerate}
    \item \textbf{Structural Hamming Distance} (SHD) measures the graph edit distance between the predicted DAG and the ground-truth DAG, as defined in \cite{shd}.
    Lower is better (0 is best).
    The discretization threshold is set to $0.5$.
    \item \textbf{Mean Average Precision} (mAP) calculates the area under the precision-recall curve for each edge and averages it across the entire graph.
    mAP ranges from 0 to 1 (best).
\end{enumerate}

We treat error (noise) detection as an edge-level binary classification task.
Since graph priors are symmetric, we symmetrize predicted graphs and omit the diagonal for this analysis.
A ``positive'' label is an edge that is flipped in the noisy prior, while a ``positive'' prediction is an edge whose presence in the prediction differs from that in the noisy prior.
We binarize with the standard threshold of 0.5.
\begin{enumerate}
    % \item \textbf{Mean Average Precision} (mAP) computes the area under the precision-recall curve for error binary prediction on each edge and averages it across the entire graph.
    % \item \textbf{Area Under the Receiver Operating Characteristic Curve} (AUC) measures the ability of the model to distinguish between positive and negative edge classifications.
    \item \textbf{Acc} is the un-weighted accuracy of identifying errors in the undirected graph.
    \item \textbf{F1 Score} is the harmonic mean of precision and recall, which provides a unified measure that considers both false positives and negatives.
\end{enumerate}

\section{Results}

\subsection{Real experiments}

Table~\ref{table:sachs} illustrates our results on the Sachs proteomics dataset with varying graph priors.
Incorporating synthetic KGs with $10\%$ and $25\%$ noise consistently improves performance in predicting the ground truth graph for \ours{} on mAP, but the effect is less evident on SHD.
When provided with the CORUM KG, which contains a much higher noise level ($34\%$), the advantage of leveraging prior knowledge diminishes.
In terms of identifying noisy edges,
\ours{} maintains robust performance across all prior knowledge settings, while the baselines are less consistent.

% ------------------------

% Synthetic results. We do this because real datasets are very hard to evaluate, and no one knows how to evaluate biology.
% \begin{enumerate}
%     \item Full dataset available, say lots of samples
%     \item Down-sampling analysis, show that KG helps. Motivation is that in real experiments, e.g. Perturb-seq / CROP-seq, don't have that many points per perturbation compared to the dimensionality of the data
% \end{enumerate}

% Real results
% \begin{enumerate}
%     \item Sachs dataset, noisy graph from CORUM
%     \item Causalbench \citep{causalbench} dataset, noisy graph from Gene Ontology
% \end{enumerate}

\subsection{Synthetic experiments}

\begin{table*}[t]
\setlength\tabcolsep{2.8 pt}
\caption{Causal discovery results on synthetic datasets where $N=10, E=10$.
Each setting is the mean over \emph{five} distinct datasets.
Best results for each data setting are denoted in \textbf{bold}.
Each dataset contains 1000 observational samples and the specified number of samples per intervention.
The symbol
$^\dagger$ indicates that \ours{} and \textsc{Sea} were not trained on this setting (for fair comparison to \textsc{Sea}).
% The best two results of a specific dataset are in bold, where the best results are highlighted in color. The best results in baseline methods are underlined.
% Preferably not linear Gaussian 10/10 since that's empirically solved and annoying to explain why.
% Baselines are described in detail in \ref{subsec:baselines}.
% More in appendix \ref{sec:more-experiments}.
}
\vspace{-0.1in}
\label{table:synthetic10}
\begin{center}
\begin{small}
\begin{tabular}{ll l rr rr rr rr}% r}
\toprule
Samples & Noise & Model 
& \multicolumn{2}{c}{Linear} 
% & \multicolumn{2}{c}{NN add.}
& \multicolumn{2}{c}{NN non-add.}
& \multicolumn{2}{c}{Sigmoid$^\dagger$}
& \multicolumn{2}{c}{Polynomial$^\dagger$} 
% & Overall
\\
\cmidrule(l{\tabcolsep}){4-5}
\cmidrule(l{\tabcolsep}){6-7}
\cmidrule(l{\tabcolsep}){8-9}
\cmidrule(l{\tabcolsep}){10-11}
&&& \multicolumn{1}{c}{mAP $\uparrow$} & \multicolumn{1}{c}{SHD $\downarrow$} & \multicolumn{1}{c}{mAP $\uparrow$} & \multicolumn{1}{c}{SHD $\downarrow$} & \multicolumn{1}{c}{mAP $\uparrow$} & \multicolumn{1}{c}{SHD $\downarrow$} & \multicolumn{1}{c}{mAP $\uparrow$} & \multicolumn{1}{c}{SHD $\downarrow$} \\ %& \multicolumn{1}{c}{Time(s) $\downarrow$} \\
\midrule
\multirow{10}{*}{50} & \multirow{4}{*}{Vanilla} 
% & \textsc{Gies} & \ms{0.75}{0.11}& \ms{4.8}{2.60}& \ms{0.61}{0.10}& \ms{8.1}{2.55}& \ms{0.65}{0.14}& \ms{6.4}{2.80}& \ms{0.55}{0.11}& \ms{8.1}{2.98}\\
& \textsc{Dcdi-G} &\ms{0.27}{0.15} &\ms{12.7}{4.22} &\ms{0.44}{0.12} &\ms{9.1}{3.56} & \ms{0.25}{0.15}& \ms{8.9}{3.88}& \ms{0.27}{0.10}&\ms{8.5}{1.36}   \\
&& \textsc{Dcdi-Dsf} & \ms{0.26}{0.09}&\ms{17.6}{5.08} &\ms{0.16}{0.04} &\ms{27.6}{4.34} &\ms{0.13}{0.05} &\ms{26.8}{5.23} &\ms{0.11}{0.03} &\ms{28.7}{3.82}  \\
&& \textsc{Bacadi} &\ms{0.42}{0.23} &\ms{11.8}{4.49} &\ms{0.39}{0.14} &\ms{14.3}{4.50} &\ms{0.39}{0.18} &\ms{11.0}{4.15} &\ms{0.28}{0.11} &\ms{12.4}{2.62}  \\
&& \textsc{Sea} & \ms{0.94}{0.05}& \ms{2.0}{1.55}& \ms{0.84}{0.07}& \ms{5.4}{1.43}& \ms{0.78}{0.18}& \ms{4.2}{2.36}& \ms{0.64}{0.15}& \mstop{6.2}{2.48}\\

\cmidrule(l{\tabcolsep}){2-11}

% && \textsc{Gies} & \ms{0.71}{0.13}& \ms{5.1}{1.37}& \ms{0.55}{0.11}& \ms{9.1}{2.66}& \ms{0.63}{0.14}& \ms{7.5}{3.72}& \ms{0.55}{0.11}& \ms{8.2}{2.96}\\
% & \multirow{3}{*}{$p=0.25$} & \textsc{Dcdi-G} & \ms{0.36}{0.13} & \ms{6.1}{1.45}& \ms{0.58}{0.20}&\ms{6.0}{3.69} &\ms{0.34}{0.18} &\ms{6.9}{1.92} &\ms{0.33}{0.12} &\ms{6.8}{1.72}  \\
% && \textsc{Dcdi-Dsf} & \ms{0.38}{0.17}&\ms{8.5}{4.15} &\ms{0.23}{0.05} &\ms{16.8}{2.64} &\ms{0.18}{0.09} & \ms{15.4}{5.61}&\ms{0.13}{0.03} &\ms{15.4}{4.10}  \\
% && \ours{} & \ms{0.93}{0.07}& \ms{1.9}{1.51}& \ms{0.84}{0.08}& \ms{6.4}{2.73}& \ms{0.80}{0.14}& \ms{4.5}{2.46}& \ms{0.64}{0.12}& \ms{6.5}{2.62}\\

% && \textsc{Gies} & \ms{0.71}{0.13}& \ms{5.1}{1.37}& \ms{0.55}{0.11}& \ms{9.1}{2.66}& \ms{0.63}{0.14}& \ms{7.5}{3.72}& \ms{0.55}{0.11}& \ms{8.2}{2.96}\\
& \multirow{3}{*}{$p=0.10$} & \textsc{Dcdi-G} & \ms{0.45}{0.14} &\ms{4.8}{1.17} & \ms{0.69}{0.13} & \mstop{3.7}{2.00} & \ms{0.36}{0.25} & \ms{6.7}{3.23} & \ms{0.35}{0.14}& \ms{6.5}{2.33} \\
&& \textsc{Dcdi-Dsf} & \ms{0.41}{0.16} & \ms{7.0}{1.84}&\ms{0.42}{0.11} &\ms{7.8}{2.09} & \ms{0.28}{0.18} & \ms{9.6}{3.44}& \ms{0.18}{0.11} & \ms{11.1}{3.70} \\
&& \ours{} & \mstop{0.95}{0.05}& \mstop{1.7}{1.19}& \mstop{0.88}{0.06}& \ms{4.1}{1.70}& \mstop{0.83}{0.17}& \mstop{3.6}{2.65}& \mstop{0.67}{0.12}& \mstop{6.2}{2.44}\\

\cmidrule(l{\tabcolsep}){2-11}

% && \textsc{Gies} & \ms{0.71}{0.13}& \ms{5.1}{1.37}& \ms{0.55}{0.11}& \ms{9.1}{2.66}& \ms{0.63}{0.14}& \ms{7.5}{3.72}& \ms{0.55}{0.11}& \ms{8.2}{2.96}\\
& \multirow{3}{*}{$p=0.25$} & \textsc{Dcdi-G} & \ms{0.36}{0.13} & \ms{6.1}{1.45}& \ms{0.58}{0.20}&\ms{6.0}{3.69} &\ms{0.34}{0.18} &\ms{6.9}{1.92} &\ms{0.33}{0.12} &\ms{6.8}{1.72}  \\
&& \textsc{Dcdi-Dsf} & \ms{0.38}{0.17}&\ms{8.5}{4.15} &\ms{0.23}{0.05} &\ms{16.8}{2.64} &\ms{0.18}{0.09} & \ms{15.4}{5.61}&\ms{0.13}{0.03} &\ms{15.4}{4.10}  \\
&& \ours{} & \ms{0.93}{0.07}& \ms{1.9}{1.51}& \ms{0.84}{0.08}& \ms{6.4}{2.73}& \ms{0.80}{0.14}& \ms{4.5}{2.46}& \ms{0.64}{0.12}& \ms{6.5}{2.62}\\

\midrule
% \midrule

\multirow{10}{*}{100} & \multirow{4}{*}{Vanilla}
% & \textsc{Gies} & \ms{0.74}{0.10}& \ms{4.4}{2.20}& \ms{0.62}{0.10}& \ms{8.3}{2.65}& \ms{0.69}{0.12}& \ms{7.0}{2.93}& \ms{0.59}{0.10}& \ms{7.1}{2.81}\\
& \textsc{Dcdi-G} 
&\ms{0.48}{0.15} &\ms{4.3}{1.27} & \ms{0.56}{0.12}&\ms{5.9}{2.21} &\ms{0.34}{0.19} &\ms{7.1}{3.36} &\ms{0.39}{0.16} &\ms{6.6}{2.11} \\
&& \textsc{Dcdi-Dsf} & \ms{0.22}{0.05}&\ms{21.2}{4.85} & \ms{0.19}{0.05}&\ms{25.9}{3.36} &\ms{0.17}{0.09} & \ms{24.4}{5.71}& \ms{0.10}{0.03}&\ms{28.3}{4.27}  \\
&& \textsc{Bacadi} &\ms{0.36}{0.23} &\ms{11.8}{3.68} &\ms{0.40}{0.11} &\ms{14.5}{3.93} &\ms{0.38}{0.15} &\ms{11.1}{3.81} &\ms{0.27}{0.12} &\ms{13.3}{3.38}  \\
&& \textsc{Sea} & \ms{0.93}{0.06}& \ms{2.2}{1.25}& \ms{0.84}{0.07}& \ms{5.5}{2.33}& \ms{0.84}{0.16}& \ms{3.8}{2.32}& \ms{0.63}{0.15}& \mstop{6.1}{2.66}\\

\cmidrule(l{\tabcolsep}){2-11}
& \multirow{3}{*}{$p=0.10$}
% & \textsc{Gies} & \ms{0.72}{0.11}& \ms{5.0}{2.05}& \ms{0.60}{0.12}& \ms{7.8}{3.28}& \ms{0.71}{0.10}& \ms{6.3}{2.79}& \ms{0.59}{0.09}& \ms{7.6}{3.20}\\
&\textsc{Dcdi-G} & \ms{0.48}{0.18}& \ms{4.2}{1.40} & \ms{0.63}{0.17} & \ms{4.5}{2.66} & \ms{0.39}{0.22} & \ms{6.4}{2.94}& \ms{0.38}{0.15} & \ms{6.3}{1.49}  \\
&& \textsc{Dcdi-Dsf} & \ms{0.41}{0.21} & \ms{7.4}{3.83}& \ms{0.45}{0.21} & \ms{8.6}{4.29}& \ms{0.23}{0.14}& \ms{10.1}{3.08}& \ms{0.21}{0.10} & \ms{9.5}{1.91}  \\
&& \ours{} & \ms{0.94}{0.06}& \ms{1.8}{1.17}& \mstop{0.90}{0.05}& \mstop{4.3}{1.79}& \mstop{0.85}{0.15}& \mstop{3.3}{2.41}& \mstop{0.69}{0.17}& \ms{6.2}{2.44}\\

\cmidrule(l{\tabcolsep}){2-11}
& \multirow{3}{*}{$p=0.25$}
% & \textsc{Gies} & \ms{0.72}{0.11}& \ms{5.0}{2.05}& \ms{0.60}{0.12}& \ms{7.8}{3.28}& \ms{0.71}{0.10}& \ms{6.3}{2.79}& \ms{0.59}{0.09}& \ms{7.6}{3.20}\\
&\textsc{Dcdi-G} & \ms{0.45}{0.17}& \ms{4.9}{1.37} & \ms{0.66}{0.14} & \ms{4.5}{3.11} & \ms{0.31}{0.24} & \ms{7.8}{4.42} &\ms{0.34}{0.20} & \ms{6.8}{1.66}  \\
&& \textsc{Dcdi-Dsf} & \ms{0.31}{0.17}&\ms{11.1}{4.64} &\ms{0.27}{0.07} &\ms{14.0}{3.52} & \ms{0.19}{0.11} & \ms{17.0}{4.38}& \ms{0.14}{0.04}& \ms{14.9}{3.62} \\
&& \ours{} & \mstop{0.95}{0.05}& \mstop{1.5}{0.92}& \ms{0.86}{0.05}& \ms{4.9}{1.81}& \ms{0.82}{0.15}& \ms{4.1}{2.26}& \ms{0.67}{0.16}& \ms{6.3}{2.69}\\

% \midrule\midrule
% 10\% noise& \textsc{Dcdi-G} & \ms{0.45}{0.14} &\ms{4.8}{1.17} & \ms{0.69}{0.13} & \ms{3.7}{2.00} & \ms{0.36}{0.25} & \mstop{6.7}{3.23} & \ms{0.35}{0.14}& \ms{6.5}{2.33} \\
% % 50 samples
% & \textsc{Dcdi-Dsf} & \ms{0.41}{0.16} & \ms{7.0}{1.84}&\ms{0.42}{0.11} &\ms{7.8}{2.09} & \ms{0.28}{0.18} & \ms{9.6}{3.44}& \ms{0.18}{0.11} & \ms{11.1}{3.70} \\
% % & \textsc{Bacadi} & & & & & & & &  \\
% \cmidrule(l{\tabcolsep}){3-10}%{3-11}
% & \ours{} &
% \mstop{0.86}{0.12}&\mstop{2.2}{1.83}&
% \mstop{0.92}{0.11}&\mstop{2.7}{1.55}&
% \mstop{0.76}{0.13}&\mstop{4.5}{2.01}&
% \ms{0.63}{0.13}&\ms{6.9}{3.14}
% \\
% & \ours{} & \ms{0.95}{0.05}& \ms{1.8}{1.33}& \ms{0.91}{0.07}& \ms{3.0}{1.95}& \ms{0.83}{0.15}& \ms{3.1}{1.84}& \ms{0.77}{0.16}& \ms{4.2}{2.40}\\

% \midrule\midrule 
% 10\% noise& \textsc{Dcdi-G} & \ms{0.48}{0.18}& \ms{4.2}{1.40} & \ms{0.63}{0.17} & \ms{4.5}{2.66} & \ms{0.39}{0.22} & \ms{6.4}{2.94}& \ms{0.38}{0.15} & \ms{6.3}{1.49}  \\
% 100 samples& \textsc{Dcdi-Dsf} & \ms{0.41}{0.21} & \ms{7.4}{3.83}& \ms{0.45}{0.21} & \ms{8.6}{4.29}& \ms{0.23}{0.14}& \ms{10.1}{3.08}& \ms{0.21}{0.10} & \ms{9.5}{1.91}  \\
% % & \textsc{Bacadi} & & & & & & & &  \\
% \cmidrule(l{\tabcolsep}){3-10}%{3-11}
% & \ours{} &
% \ms{0.84}{0.09}&\ms{2.9}{1.58}&
% \ms{0.90}{0.07}&\ms{3.1}{1.87}&
% \ms{0.72}{0.12}&\ms{4.9}{2.59}&
% \mstop{0.71}{0.11}&\mstop{6.0}{2.32}
% \\
\bottomrule
\end{tabular}
% \end{sc}
\end{small}
\end{center}
\vspace{-0.1in}
\end{table*}

% on N=20, E=20, SEA performs too well (0.9x mAP vs. 0.78 ours) so I'm replacing this table with ablations 哈哈哈

\begin{figure}[t]
    \centering
    \includegraphics[width=\linewidth]{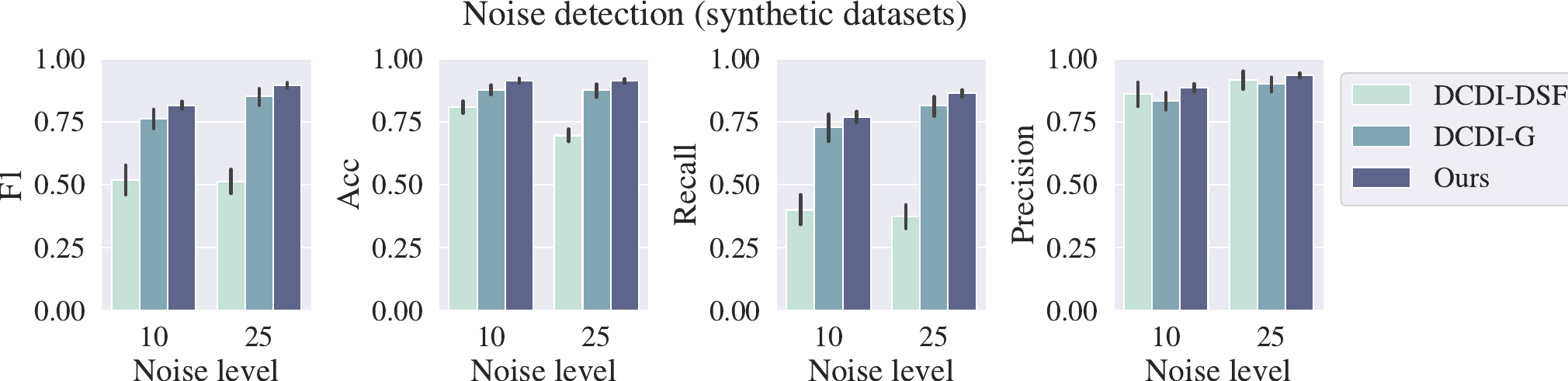}
    \caption{Noise detection on synthetic datasets.
    \ours{} outperforms \textsc{Dcdi} at identifying errors in the graph prior, for both noise levels.}
    \label{fig:noise}
\end{figure}

Given the limited availability and challenges of evaluating real-world datasets -- particularly in biological contexts where no standard evaluation exists -- we perform a comprehensive comparison across synthetic datasets under various settings to assess the performance of each model (Table~\ref{table:synthetic10}, Figure~\ref{fig:noise}).
% and \ref{table:synthetic20}).
We make the following observations.

% Each setting, as in the first column, comprises a prior knowledge configuration: Vanilla, prior noisy graphs with $25\%$ noise, and prior noisy graphs with $10\%$ noise. Here, "Vanilla" refers to the default setting for each model, meaning that no prior knowledge is provided to \textsc{Dcdi-G} and \textsc{Dcdi-Dsf}, while the prior for \textsc{BaCaDi} follows their default setting. Additionally, we apply down-sampling to the number of data points in each intervention regime, using either $50$ or $100$ samples per intervention.

% \subsubsection{KGs do help improve performance}
\paragraph{Graph priors are useful in low-data scenarios, even at higher levels of noise.}

Incorporating graph priors improves performance for both our model and the \textsc{Dcdi} baselines.
For \textsc{Dcdi-G} and \textsc{Dcdi-Dsf}, the performance with a prior noisy graph containing $25\%$ noise is significantly better than that of the vanilla setting, particularly in the down-sampling scenario with only $50$ samples per interventional regime.
Our model is more sensitive to noise (perhaps because the \textsc{Sea} baseline starts at a high level), but a graph prior with $10\%$ is consistently helpful across both data settings.
% Moreover, providing a noisy graph with $10\%$ noise yields even better results than the $25\%$ noise setting. 
% Starting the optimization process with prior knowledge graphs (KGs) offers a significant advantage by guiding the algorithm toward a more optimal solution. Unlike the default initialization, where the weighted adjacency matrix is fully connected (except for the diagonal), KGs help direct the optimization of the Gumbel adjacency matrix, mitigating the risk of getting trapped in local optima. This leads to a more efficient optimization process and better overall performance, as demonstrated by the superior results when noisy graphs with $10\%$ and $25\%$ noise were used in \textsc{Dcdi-G} and \textsc{Dcdi-DSF}.

\paragraph{Graph priors are particularly helpful on hard settings.}

On easy settings (e.g. linear), the \textsc{Sea} baseline already achieves near-perfect mAP and SHD scores, as the assumptions of \textsc{Sea}'s summary statistics (inverse covariance) match the linear Gaussian setting exactly.
Correspondingly, our approach's improvement is less pronounced on linear Gaussian.
However, there is slight mismatch between \textsc{Sea}'s assumptions and the remaining three settings; and as expected, providing a graph prior on those cases leads to larger improvements.

\begin{figure}[t]
    \centering
    \includegraphics[width=0.5\linewidth]{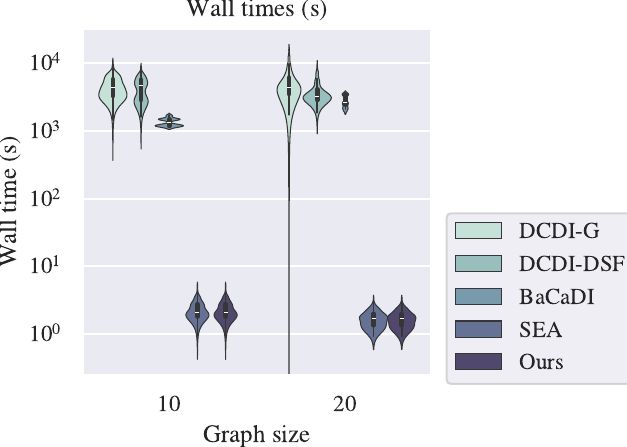}
    \caption{Runtime analysis of all algorithms.
    Amortized inference approaches (\textsc{Sea}, \textsc{Ours}) are orders of magnitude faster.}
    \label{fig:runtime}
\end{figure}

\paragraph{Additional observations}

\ours{} outperforms \textsc{Dcdi} at noise detection across noise levels (Figure~\ref{fig:noise}).
As an amortized inference method, \ours{} also achieves runtimes that are orders of magnitude faster than \textsc{Dcdi} (Figure~\ref{fig:runtime}).
The inclusion of a graph prior does not negatively impact runtime, compared to \textsc{Sea}.

\section{Conclusion}

In this work, we have presented an amortized inference algorithm for refining prior knowledge into data-dependent graphs.
We demonstrated in synthetic and real settings that incorporating prior knowledge is particularly helpful in low-data settings, and that our approach is able to detect errors in these priors with high accuracy.
However, we also observed that biological knowledge graphs contain high levels of noise in their connectivity alone, so it could be valuable to incorporate semantic information regarding the graphs and/or data for future work.

\section*{Acknowledgements}

This material is based upon work supported by the National Science Foundation Graduate Research Fellowship under Grant No. 1745302. We would like to acknowledge support from the NSF Expeditions grant (award 1918839: Collaborative Research: Understanding the World Through Code), Machine Learning for Pharmaceutical Discovery and Synthesis (MLPDS) consortium, and the Abdul Latif Jameel Clinic for Machine Learning in Health.

\newpage
\bibliography{references}
\bibliographystyle{iclr}

\end{document}